\documentclass[amsmath, amsfonts, showpacs, twocolumn, prl]{revtex4}
\usepackage{graphicx}
\usepackage{epsfig}
\usepackage{bm}
\usepackage{dcolumn}
\usepackage{amsmath}
\usepackage{amssymb}

\begin{document}

\title{Nonequilibrium Spectroscopy of Topological Edge Liquids}

\author{Stanislav S. Apostolov}
\affiliation{Department of Physics and Astronomy, Michigan State
University, East Lansing, Michigan 48824, USA}

\author{Alex Levchenko}
\affiliation{Department of Physics and Astronomy, Michigan State
University, East Lansing, Michigan 48824, USA}

\begin{abstract}
We develop a theory for energy and spatially resolved tunneling
spectroscopy of topological quantum spin Hall helical states driven
out of equilibrium. When a helical liquid is constrained between two
superconducting reservoirs transport at the edge is governed by
multiple Andreev reflections. The resulting quasiparticle
distribution functions of the edge channels exhibit multiple
discontinuities at subgap energies with the periodicity of an
applied voltage. The combined effect of interactions, disorder, and
normal scattering off the superconducting interface leads to the
inelastic processes mixing different helicity modes, thus causing
smearing of these singularities. If equilibration is strong, then the
distribution functions of the edge channels tend to collapse into a
Fermi-like function with an effective temperature determined by the
superconducting gap, applied voltage, and intraedge interaction
parameter. We conclude that mapping out nonequilibrium distribution
functions may help to quantify the relative importance of various
relevant perturbations that spoil ideally ballistic edge transport.
\end{abstract}

\date{May 19, 2014}


\maketitle

Energy and spatially resolved tunneling experiments provide
extremely sensitive spectroscopic tools for accessing and
characterizing nonequilibrium states. This technique was introduced
in the context of mesoscopic quantum
wires~\cite{Birge-PRL97,Pierre-PRL01,Anthore-PRL03} and later
successfully applied to probe quasiparticle distributions in carbon
nanotubes~\cite{Mason-PRL09} and the edge channels of the integer
quantum Hall effect~\cite{Altimiras-NP10}. The power of such
experiments is that they allow one to deduce the scaling of
quasiparticle lifetimes on energy and also to elucidate the
microscopic scattering processes responsible for the relaxation.
Motivated by these experimental capabilities, we develop a theory
for the out-of-equilibrium energy-resolved spectroscopy of a new
class of one-dimensional liquids as realized in the quantum spin
Hall (QSH) insulators~\cite{Kane-Mele,Bernevig-SCZ}.

A key signature of the QSH effect is the appearance of gapless edge
states coexisting with gapped bulk states. These edge modes
counterpropagate and carry opposite spins. Such correlation between
the direction of motion and spin orientation facilitates the term
helical liquid~\cite{Wu-PRL06}. Owing to protection by time-reversal
symmetry, helical modes are immune to single-particle elastic
backscattering by nonmagnetic disorder, thereby providing
opportunities for dissipationless transport with a universally
quantized temperature-independent conductance of $e^2/h$ per helical
edge. This expectation has been confirmed experimentally in the
topologically nontrivial phase of HgTe/HgCdTe and InAs/GaSb
heterostructures hosting helical edge states with observed
conductance close to the quantum
value~\cite{Konig-Science07,RRD-PRL11}. However, clear deviations
from this limit have also been identified: the conductance is
suppressed for increasing system
size~\cite{Roth-Science09,Gusev-PRB11,Konig-PRX13,Nowack-NM13,Suzuki-PRB13},
thus implying the presence of inelastic relaxation. These findings
have attracted a great deal of theoretical attention and triggered
multiple proposals for possible scattering mechanisms affecting the
ideally ballistic edge
transport~\cite{Kane-Mele,Wu-PRL06,Moore-HES,Maciejko-PRL09,Tanaka-PRL11,Trauzettel,Kharitonov,Strom,Lezmy,
Schmidt-PRL12,Vayrynen-PRL13,Eriksson-PRB13}. We briefly summarize
various scenarios for the equilibration processes and temperature
scaling of the corresponding scattering rates.

\begin{figure}
  \includegraphics[width=9cm]{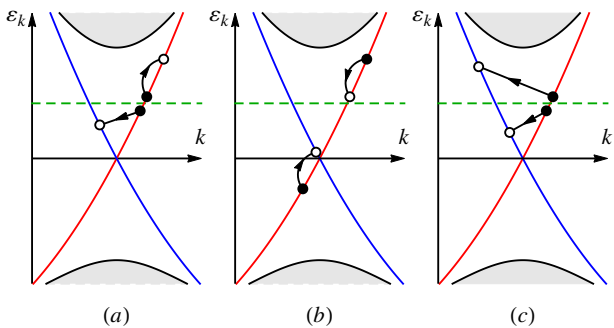}\\
 \caption{An allowed backscattering processes: (a) single-particle backscattering
 accompanied by particle-hole excitation. At weak interactions $K\simeq1$ the rate of
 this process scales with temperature as $T^4$. (b) Umklapp scattering, which goes with
 a rate $\propto T^5$ if the Fermi energy is tuned to the time-reversal invariant point $T\gg E_F$,
 while this rate is exponentially suppressed as $\propto e^{-E_F/T}$ in the opposite limit $T\ll E_F$.
 (c) Impurity-assisted inelastic two-particle backscattering that scales with the
 temperature as $T^6$. }\label{Fig-1}
\end{figure}

The most obvious reason for electron backscattering is magnetic
disorder that can flip the spin and thus induce transitions between
counterpropagating edge modes. The rate of this process follows a
power-law temperature dependence $T^{2K-2}$ at low temperatures with
the exponent determined by the Luttinger liquid interaction
parameter
$K$~\cite{Wu-PRL06,Maciejko-PRL09,Tanaka-PRL11,Eriksson-PRB13}. The
same temperature scaling of the current backscattered from an
impurity applies to the case when time-reversal symmetry is broken
directly by an applied magnetic field. If time-reversal symmetry is
preserved then the leading-order inelastic processes involve
backscattering of electron
pairs~\cite{Kane-Mele,Wu-PRL06,Moore-HES,Trauzettel,Lezmy,Eriksson-PRB13}.
The latter are either umklapp or disorder-mediated processes due to
momentum nonconservation. Umklapp scattering has an exponentially
slow rate $e^{-E_F/T}$ unless the Fermi energy $E_F$ is tuned to the
immediate vicinity of the Dirac point. Impurity-assisted scattering
is less susceptible to the position of the Fermi level, and the
corresponding rate scales as $T^{4K}$ or $T^{8K-2}$ depending on
whether $K\gtrless1/2$. The above arguments apply to helical liquids
that possess an additional axial symmetry. It has been recently
emphasized~\cite{Strom,Schmidt-PRL12,Rothe-NJP10,Virtanen-PRB12}
that such an auxiliary symmetry, if present in the system, could be
easily lifted by the combined effect of a strong spin-orbit
interaction and back-gate-induced electric field which causes the
spin quantization axis to rotate with the momentum. This mechanism
opens a new channel for inelastic scattering, namely single-particle
backscattering accompanied by particle-hole excitation, which also
requires impurities to remove constraints imposed by momentum
conservation. The rate of this process scales with energy as
$T^{2K+2}$. Interestingly, the same mechanism also allows
single-particle backscattering even for the translationally
invariant edge, however with a slower rate due to kinematic
constraints. Figure~\ref{Fig-1} summarizes the leading inelastic
two-particle processes for the time-reversal-symmetric QSH edge.

The value of the Luttinger liquid interaction parameter $K$ within
the edge states depends on the details of the heterostructure and
materials. It can be estimated as $K=[1+r_s\ln(d/a)]^{-1/2}$, where
$r_s=2e^2/\pi^2v_F\kappa$ is the electron gas parameter, $d$ is the
distance to a nearby metallic gate, and $a$ is the short-distance
cutoff for the Coulomb interaction, which is either determined by
the microscopic screening length or the thickness of the quantum
well. For typical values of the Fermi velocity $v_F\simeq5\times
10^5$ m/s and high dielectric constant $\kappa\simeq12$ of HgTe, one
estimates $K>0.9$ for values of $\ln(d/a)<3$. In the following we
concentrate on the weakly interacting limit assuming $K\simeq1$.

Given the plethora of scattering events, it is desirable to have an
experimental probe that may potentially distinguish between
different possibilities or perhaps elucidate their relative
importance. As alluded to at the beginning, such a probe may be
provided by energy-resolved spectroscopy of the helical edge
channels, and our motivation is to develop corresponding theory. We
propose to consider a QSH edge that is constrained between two
ordinary $s$-wave superconductors and driven out of equilibrium by
an applied voltage $V$ between them. We assume that due to the
proximity effect the superconductors induce a gap $\Delta$ over a
finite length $l$ of the helical liquid. The latter is determined by
the lithographic thickness of the superconducting leads, which we
assume to be up to a few times larger than the superconducting
coherence length $l\gtrsim \xi=v_F/\Delta$. The distance $L$ between
the superconductors is assumed to be in the limit $L\gg\{l,\xi\}$.
We are interested in calculating the energy and spatially dependent
profiles of the distribution functions $f^{\pm}_\varepsilon(x)$ for
the right- and left-moving $(\pm)$ helicity channels, neither of
which is assumed to be approximated by the equilibrium Fermi
function. We demonstrate that $f^{\pm}_\varepsilon(x)$ is extremely
sensitive to the interaction effects and relevant perturbations
affecting ideally ballistic transport. The choice in favor of the
above specified geometry is twofold. First, at small voltages
$eV<\Delta$ the transport along the edge is governed by multiple
Andreev reflections (MAR), which strongly magnify the structure of
the distribution functions $f^{\pm}_\varepsilon(x)$ at subgap
energies, thus making it easily detectable experimentally. Second,
due to MAR trajectories quasiparticles traverse the edge between
superconductors many times, which effectively increases the
probability of inelastic scattering even for relatively weak
interactions. It should be stressed that MAR have been recently
observed in the QSH regime of InAs/GaSb quantum
wells~\cite{RRD-PRL12} and in InSb/InAs
nanowires~\cite{Nilsson-NanoLett12}. Induced superconductivity has
also been achieved in QSH channels of HgTe/HgCdTe
heterostructures~\cite{Molenkamp-Yacoby}. Our focus on inelastic
scattering complements other recent interesting proposals addressing
nonequilibrium transport and proximity-effect-related phenomena with
QSH
edges~\cite{Virtanen-PRB12,Fu-Kane,Cayssol-PRB10,Houzet-PRL11,Garate-PRB12,Ilan-PRL12,Beenakker-PRL13}.
Furthermore, our work may also be of special interest in view of
ongoing efforts in studying equilibration of one-dimensional liquids
beyond the Luttinger liquid paradigm where inelastic multiparticle
scattering processes play a central
role~\cite{Barak-NP10,Karzig-PRL10,Gangardt-PRL10,Micklitz-PRL11,Matveev-PRB12,Lamacraft-PRA13}.

The method of calculating the distribution functions that we employ
in this study is based on the quantum kinetic equation
\begin{equation}\label{Kin-Eq}
\pm\partial_xf^\pm_\varepsilon(x)=-\mathrm{St}\{f^\pm_\varepsilon(x)\}.
\end{equation}
The same technique has been previously used to describe ballistic
and diffusive superconductor--normal-metal wire--superconductor (SNS)
quasi-one-dimensional proximity circuits in the regime of
MAR~\cite{OTBK,Nagaev}. The collision integral
$\mathrm{St}\{f^\pm_\varepsilon\}$ of the kinetic equation
\eqref{Kin-Eq} depends on a particular type of scattering affecting
the distribution function. Although all the backscattering processes
shown in Fig.~\ref{Fig-1} appear to be of the same order in
interaction (at least in the weak-coupling limit), we choose
inelastic single-particle backscattering [Fig.~\ref{Fig-1}(a)] as a
guiding example. The reason for this is based on the observation
that this process has the lowest scaling with energy, which implies
the fastest relaxation time. Furthermore, this scattering does not
assume axial spin symmetry, which is expected to be a generic
feature of helical states. The corresponding collision integral
reads
\begin{eqnarray}\label{St}
&&\mathrm{St}\{f^\alpha_\varepsilon\}=-\lambda\sum_{\beta\sigma\varsigma=-\alpha}\iint
d\varepsilon_1 d\varepsilon_2
M^{\alpha\beta}_{\varepsilon,\varepsilon_1,\varepsilon_2}\nonumber
\\
&&[f^\alpha_\varepsilon
f^\beta_{\varepsilon_1-\varepsilon_2}g^\sigma_{\varepsilon-\varepsilon_2}g^\varsigma_{\varepsilon_1}-
g^\alpha_\varepsilon
g^\beta_{\varepsilon_1-\varepsilon_2}f^\sigma_{\varepsilon-\varepsilon_2}f^\varsigma_{\varepsilon_1}],
\end{eqnarray}
where summation over the helicity indices $\beta, \sigma$ and
$\varsigma$ goes in such a way that their product equals $-\alpha$.
The kernel of the collision integral is quadratic in energy
$M^{\alpha\beta}_{\varepsilon,\varepsilon_1,\varepsilon_2}=
(\varepsilon-\varepsilon_1+\alpha\beta\varepsilon_2)^2$, and we also
introduced the notation $g^\pm_\varepsilon=1-f^\pm_\varepsilon$. In
Eq.~\eqref{Fig-1} we rescaled the coordinate $x\rightarrow x/L$,
while in Eq.~\eqref{St} we rescaled all energies in units of the
superconducting gap, $\varepsilon\rightarrow\varepsilon/\Delta$, so
that the interaction parameter $\lambda$ is explicitly
dimensionless. Its precise value depends on the model. For example,
for the case of sharp disorder it can be derived perturbatively to
second order in electron-electron and electron-impurity scattering
with the result
$\lambda\simeq[(U_0+U_{2k_F})W_{2k_F}/4v^2_F]^2(\Delta/k_{0}v_F)^4$,
where $U_0, U_{2k_F}$, and $W_{2k_F}$ are the Fourier components of
the Coulomb and impurity potentials, respectively, while $k_0$
parametrizes the scale on which the spin quantization axis rotates
with momentum~\cite{Schmidt-PRL12}. In our analysis we keep
$\lambda$ as a small adjustable parameter. It is worth noting that
the same collision integral as in Eq.~\eqref{St} appears in a
different model of disorder~\cite{Vayrynen-PRL13}, namely,
spontaneously formed quantum dots representing the large-scale
inhomogeneities of electron and hole puddles.

\begin{figure}
  \includegraphics[width=8cm]{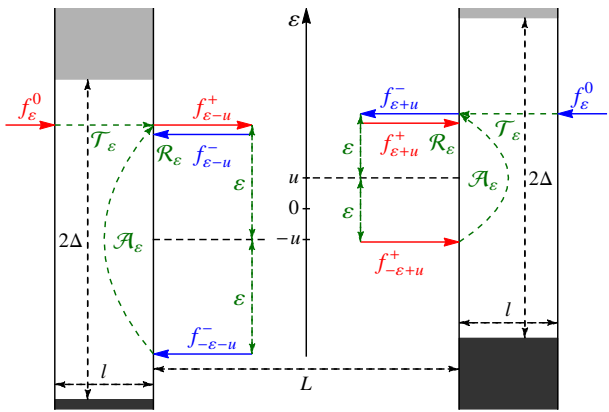}\\
  \caption{The scheme of energy bands for multiple Andreev reflection trajectories.
  Horizontal dashed lines at $\pm u$ represent Fermi levels of superconductors, $u=eV/2\Delta$.
  Lines with arrows represent incoming and outgoing particles reflected via normal
  or Andreev processes with corresponding coefficients and occupation factors indicated
  on the diagram.}\label{Fig-0}
\end{figure}

The kinetic equation should be supplemented by appropriate boundary
conditions connecting $f^\pm_\varepsilon$ to the distribution
functions in the reservoirs $f^0_\varepsilon$, which are simply the
Fermi functions. This can be accomplished by the following relations
at the interfaces
\begin{equation}\label{Bounday}
f^\mp_{\varepsilon\pm
u}=\mathcal{A}_\varepsilon(1-f^\pm_{-\varepsilon\pm
u})+\mathcal{R}_\varepsilon f^\pm_{\varepsilon\pm
u}+\mathcal{T}_\varepsilon f^0_\varepsilon,\quad x=\pm1/2,
\end{equation}
with $u=eV/2\Delta$. For technical convenience in further numerical
calculations, we choose to measure the excitation energy with
respect to a point halfway between the electrochemical potentials in
the superconductors. The corresponding energy diagram connected to
the boundary conditions \eqref{Bounday} and MAR trajectories is
shown in Fig.~\ref{Fig-0}. The first term on the right-hand side of
Eq.~\eqref{Bounday} corresponds to the Andreev process which is
based on the fact that $1-f^\pm$ accounts for the occupation
probability that an incident hole with energy $-\varepsilon$ will
reflect with amplitude probability $\mathcal{A}_\varepsilon$ and
emerge as an electron with energy $\varepsilon$. The other two terms
correspond to normal reflection and transmission with respective
probabilities $\mathcal{R}_\varepsilon$ and
$\mathcal{T}_\varepsilon$. It is also expected that for energies
deep below the superconducting gap $\varepsilon\ll-1$, the
distribution functions saturate to unity
$f^\pm_\varepsilon\rightarrow1$ since such states are fully
occupied, whereas $f^\pm_\varepsilon\rightarrow0$ in the opposite
case $\varepsilon\gg1$ for the highly excited states which are empty
(recall that $\Delta\to1$ in the rescaled energy units).

\begin{figure}
  \includegraphics[width=7cm]{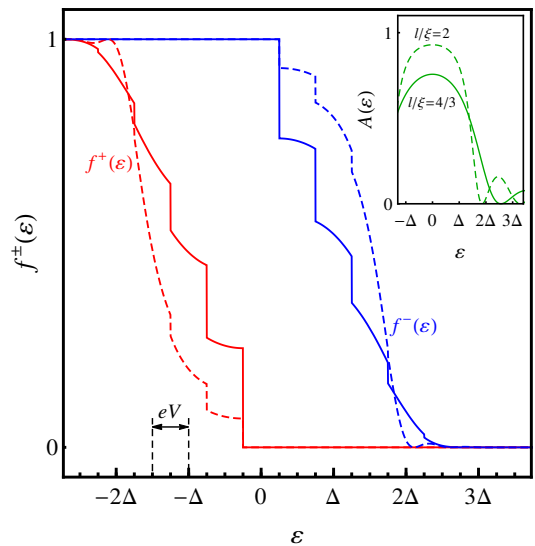}\\
  \caption{Nonequilibrium distribution functions $f^\pm_\varepsilon$ in the absence
  of inelastic scattering for $l/\xi=4/3$ (solid line) and $l/\xi=2$ (dashed line) and $eV=\Delta/2$. The inset represents energy dependence of the Andreev reflection coefficient from Eq.~\eqref{A} for the
  same choice of ratios $l/\xi$.}\label{Fig-2}
\end{figure}

The coefficient of Andreev reflection can be found by solving the
auxiliary problem of one-dimensional Dirac fermions impinging on the
potential step created by a superconducting strip of width $l$ with
the result
\begin{equation}\label{A}
\mathcal{A}_\varepsilon=\frac{1}{\varkappa^2_\varepsilon\coth^2(l\varkappa_\varepsilon/\xi)+\varepsilon^2},
\end{equation}
where $ \varkappa_\varepsilon=\sqrt{1-\varepsilon^2}$. For energies
outside of the gap $\varkappa_\varepsilon$ becomes imaginary and
$\mathcal{A}_\varepsilon$ undergoes oscillatory decay. One should
notice here that for subgap energies and a wide superconductor,
$l\gg\xi$, the Andreev coefficient $\mathcal{A}_\varepsilon$ is
nearly unity. For a narrower superconductor with $l\sim\xi$ the
evanescent Bogoliubov quasiparticle wave survives and thus reduces
$\mathcal{A}_\varepsilon$ in absolute value. This behavior is
illustrated in the inset of Fig.~\ref{Fig-2}.

For ideal edge channels with an additional spin axial symmetry the
normal reflection coefficient $\mathcal{R}_\varepsilon$ is
identically zero. Indeed, in the course of normal reflection an
electron changes its direction of propagation but conserves its
spin, which is not possible due to the helicity constraint in an
ideal QSH edge state. However, for a generic helical liquid one
expects to have a finite probability of normal reflection.
Calculation of the energy dependence of $\mathcal{R}_\varepsilon$ is
a difficult task since the result depends on the actual mechanism
that lifts spin symmetry and also on the details of the
heterostructure. We suspect that the precise energy dependence of
$\mathcal{R}_\varepsilon$ is not of principal importance because
this coefficient is small and can be treated as a perturbation. We
take in our modeling
$\mathcal{R}_\varepsilon=\gamma\mathcal{A}_\varepsilon$ with
$\gamma\ll1$ being a second control parameter of the theory. In the
case of conventional superconductor--normal-metal junctions the
parameter $\gamma$ is small in the ratio of the energy gap to the
Fermi energy $\Delta/E_F\ll1$. In the context of the QSH proximity
effect the Fermi energy is replaced by the bulk band gap of the
material $E_G$. Furthermore, $\gamma$ acquires an additional
smallness due to off-diagonal terms in spin space since the spin has
to rotate for normal reflection to occur. Based on the
$\mathbf{k}\cdot\mathbf{p}$ calculations of Ref.~\cite{Rothe-NJP10}
one may estimate $\gamma\sim
(\mathcal{E}_z/\mathcal{E}_0)(\Delta/E_G)$, where $\mathcal{E}_z$ is
the electric field generated perpendicular to the plane of a quantum
well and $\mathcal{E}_0\sim100$mV/nm. For $E_G\sim10$ meV and a
conventional superconductor, a conservative numerical estimate for
normal reflection gives $\gamma\lesssim0.1$. In the calculations we
use $\mathcal{A}_\varepsilon$ from Eq.~\eqref{A}, choose $\gamma$ as
a variable small parameter, and enforce
$\mathcal{T}_\varepsilon=1-\mathcal{A}_\varepsilon-\mathcal{R}_\varepsilon$.

\begin{figure}
  \includegraphics[width=8cm]{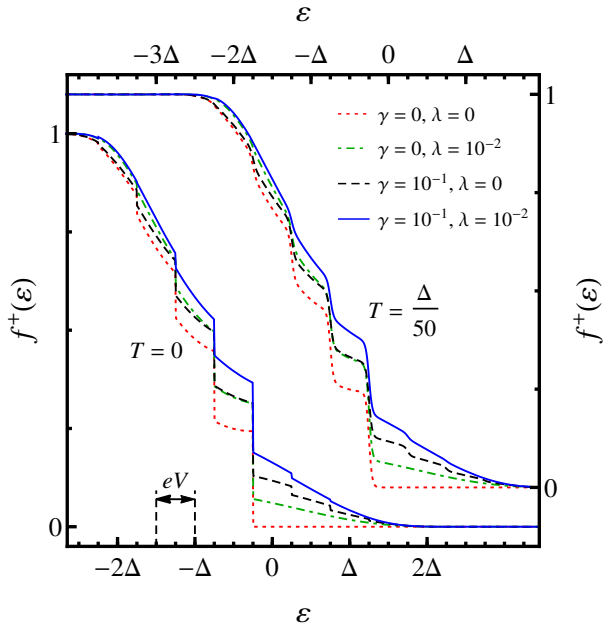}\\
 \caption{Nonequilibrium distribution functions $f^\pm_\varepsilon$
 plotted for the case $l/\xi=4/3$ and $eV=\Delta/2$ at different values of parameters
 $\gamma$ and $\lambda$ quantifying normal and inelastic scattering,
 and also at different temperatures $T=0$ and $T=\Delta/50$.
}\label{Fig-3}
\end{figure}

The nonequilibrium distribution functions $f^\pm_\varepsilon$ of an
ideal QSH edge channel can be found analytically. In the absence of
inelastic scattering the collision integral in Eq.~\eqref{St} drops
out from the kinetic problem and should be replaced by the condition
$f^\pm_\varepsilon|_{x=-1/2}=f^\pm_{\varepsilon}|_{x=1/2}$. If we
rewrite this condition in terms of the energy measured with respect
to the local value of the electrochemical potential it takes the
form $f^\pm_\varepsilon|_{x=-1/2}=f^\pm_{\varepsilon-2u}|_{x=1/2}$
merely stating that electrons with energy $\varepsilon$ at $x=-1/2$
have different energy when they arrive at the other superconductor
at $x=1/2$ due to the applied voltage bias along the edge. When
combined with the boundary conditions, Eq.~\eqref{Bounday}, one
obtains a closed set of recursion relations, which can be resolved
in the form
\begin{equation}\label{f}
f^\pm_\varepsilon=\sum^\infty_{l=0}\left[\prod^{l-1}_{m=0}
\mathcal{A}_{\varepsilon\pm(2m+1)u}\right]
\mathcal{T}_{\varepsilon\pm(2l+1)u}f^0_{\varepsilon\pm(2l+1)u}.
\end{equation}
This result is analogous to ballistic SNS channels~\cite{OTBK}.
These distributions are plotted in the main panel of
Fig.~\ref{Fig-2} for two different ratios of $l/\xi$ at zero
temperature with thus $f^0_\varepsilon=\theta_{-\varepsilon}$, where
$\theta_\varepsilon$ is the conventional Heaviside step function.
The rich subgap structure of the distribution functions is due to
MAR trajectories, and discontinuities appear in steps of the applied
voltage. It should be noted, that the case of perfect Andreev
reflection, $\mathcal{A}_\varepsilon=1$, at zero temperature
corresponds to distributions $f^\pm_\varepsilon$ that are close to
displaced Fermi functions without any special subgap features.

Next we address the role of normal and inelastic scattering on the
subharmonic gap structure of the distribution functions. This
problem was solved numerically by treating the collision integral in
the kinetic equation perturbatively by iterations. We find that a
finite value of the normal reflection coefficient leads to an
appreciable shift of the discontinuities in the distribution
function in their absolute values but not in their positions in
energy. Even more importantly, finite value of $\gamma$ results also
in the appearance of new discontinuities at energies near the
superconducting gap edges (see Fig.~\ref{Fig-3}). This leads to an
interesting conclusion that while the subgap structure of the
distribution functions is a consequence of Andreev reflection
processes, it is the normal reflection processes which sharpen the
structure. Experimentally this feature could be at least a
qualitative measure of the normal reflection in QSH devices subject
to the proximity effect. The primary role of the inelastic
scattering along the edge is to smear subgap discontinuities in the
distribution functions and also to extend the tails of these
distributions deeper into the regions below and above the
superconducting gap. This is qualitatively similar to the
finite-temperature effect, which also leads to a substantial
broadening of the subharmonic structure as illustrated in
Fig.~\ref{Fig-3}. Although the regime of strong equilibration of the
helical edges is beyond the scope of this work, we observe that at
$u\ll1$ increasing the interaction parameter to $\lambda\sim1$  is the
equivalent of having an effective temperature
$T_{\mathrm{eff}}\sim\Delta/\ln(\lambda/u^2)\ll\Delta$, and the
distribution functions tend to take a Fermi-like shape.

In summary, we have developed a kinetic equation approach for the
direct calculation of the distribution functions of the QSH helical
edges states. Our model contains the essential physical processes
leading to subharmonic gap structure, namely, Andreev and normal
reflection processes at superconducting interfaces, as well as
inelastic backscattering along the edge. We believe that our theory
will be useful for the interpretation of energy-resolved experiments
that may help to quantify the degree of intraedge equilibration,
and thus strength of interactions and the strength of normal
scattering, which is determined by the effects lifting auxiliary
spin symmetry of ideal helical edge modes.

\end{document}